\documentclass[twocolumn,superscriptaddress,letter]{revtex4}%
\usepackage{amssymb}
\usepackage{color}
\usepackage{graphicx}
\usepackage{dcolumn}
\usepackage{bm}
\usepackage[header,title,page,titletoc]{appendix}
\usepackage{amsmath}
\usepackage{subfigure}
\usepackage{amsfonts}%
\providecommand{\U}[1]{\protect \rule{.1in}{.1in}}

\begin{document}
\title{Type II Nodal line Semimetal}
\author{Jing He}
\affiliation{Department of Physics, Beijing Normal University, Beijing 100875, China}
\affiliation{Department of Physics, Hebei Normal University, Shijiazhuang 050024, China}
\author{Xiao Kong}
\affiliation{Department of Physics, Beijing Normal University, Beijing 100875, China}
\author{Wei Wang}
\affiliation{Department of Physics, Beijing Normal University, Beijing 100875, China}
\author{Su-Peng Kou}
\affiliation{Department of Physics, Beijing Normal University, Beijing 100875, China}

\begin{abstract}
Recently, topological semimetals become hot topic in condensed matter
physics, including Dirac semimetal, Weyl semimetal, and nodal line semimetal
(NLSM). In this paper, a new type of node-line semimetal - type-II NLSM is
proposed based on a two-band cubic lattice model. For type-II NLSM, the zero
energy bulk states have a closed loop in momentum space but the (local) Weyl
cones on nodal line become tilted. The effect of magnetic field and that of
correlation on type-II NLSM are studied. In particular, after considering
repulsive interaction and additional spin degrees of freedom, different types
of long range magnetic orders appear in bulk states. In addition, the
interaction-induced ferromagnetic order of surface states may exist. At
critical point between type-I NLSM and type-II NLSM, arbitrary tiny
interaction induces ferromagnetic order due to a flat band at Fermi surface.

\end{abstract}
\maketitle

\section{Introduction}

Recently, topological semimetals have attracted considerable eyes of
researchers. Compared to topological insulator, topological semimetals have
gapless bulk states and topologically protected surface Fermi arc states.
There exist different types of topological semimetals, such as Dirac semimetal
(DSM)\cite{DSM1,DSM2}, Weyl semimetal (WSM)\cite{WSM1,XWan,SM,SM1}, and nodal
line semimetal (NLSM)\cite{NLSM1,NLSM2,NLSM3,CPB}. WSM was proposed to
separate a single Dirac node into two Weyl nodes by breaking either time
reversal symmetry or inversion symmetry. The surface states of Weyl semimetal
become Fermi arc between a pair of Weyl points with opposite chiralities.
Morever, Weyl semimetals have been found in experiments such as TaAs
family\cite{Lv1,Lv2,Chang,Xu}. Nodal line semimetal is a three-dimensional
graphene-like system with low-energy relativistic excitations, but the band
touches are closed loop in momentum space instead of points. The surface
states of node-line semimetal have drumheadlike surface flat bands. The
node-line semimetal is also realized in experiments (For example Ca$_{3}%
$P$_{2}$\cite{Xie} and Cu$_{3}$PdN\cite{Yurui}).

\begin{figure}[ptbh]
\includegraphics[width = 0.5\textwidth]{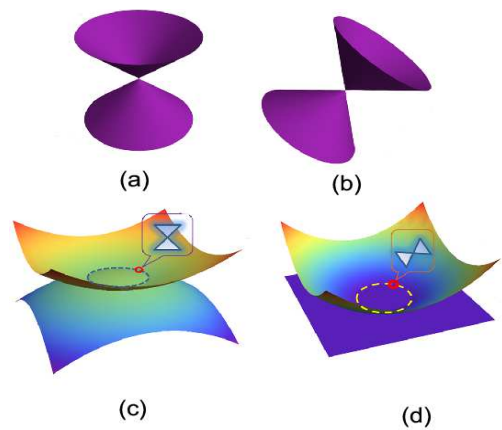}\caption{(Color online) (a)
An illustration of the low energy linear dispersion of type I Weyl semimetal.
(b) An illustration of the low energy linear dispersion of type II Weyl
semimetal, which is tilted along certain direction in Brillouin zone (BZ). The
electron and hole pockets touch, and the dispersions become anisotropic. (c)
An illustration of low energy dispersion of type-I NLSM that has a closed loop
in momentum space. The low energy effective excitation of every node on the
nodal-line is also linear. (d) An illustration of low energy dispersion of
type-II NLSM that also has a closed loop in momentum space. Due to the tilted
linear dispersion, the valence and conduction bands are asymmetry. }%
\label{fig1}%
\end{figure}

In addition, new types of WSMs are proposed which are called type-II Weyl
semimetal\cite{WSM2} and Hybrid (type-1.5) Weyl semimetal\cite{WSM3,kong}. In
these types of WSMs, Lorentz invariance of low-energy excitations is broken.
As a result, the nodes are tilted along given directions (see FIG.\ref{fig1}%
(a) and (b)) and the transport properties become anisotropic. There are many
remarkable phenomena appearing in type-II WSMs, such as the anisotropic
negative magnetoresistance effect caused by Landau level
collapsion\cite{aniso,Yao} and the existence of tilted surface
states\cite{kong}. In Hybrid (type-1.5) WSM, because the remaining symmetry
(inversion symmetry, time reversal symmetry or mirror symmetry) of two nodes
is broken, one Weyl node belongs to type-I and the other Weyl node belongs to
type-II. These new types of WSMs attracted plenty of studies in past two years.

In this paper, based on a tight-binding model, we point out that there exists
a new type of NLSM named \emph{type-II NLSM}. For type-II NLSM, the zero
energy bulk states have a closed loop in momentum space but the (local) Weyl
cones on nodal line become tilted (see FIG.\ref{fig1}(c) and (d)). In sec.II
and sec.III, we introduce a two-band tight-binding model that describes
type-II NLSM. In sec.IV, we study the effect of magnetic field on type-II NLSM
and show the Landau level collapsion in this system. Next, we study the
correlation effect on type-II NLSM and the interaction-induced magnetic order
of NLSM is found in sec.V. An interesting result is at critical point between
type-I NLSM and type-II NLSM, arbitrary tiny interaction induces ferromagnetic
order (FM) due to a flat band at Fermi surface. Finally, we give the
conclusion and propose an experimental realization in sec.VI.

\section{The nodal line Hamiltonian in real space on cubic lattice}

Firstly, we start with a nodal line semimetal from a three dimensional (3D)
tight-binding Hamiltonian on cubic lattice that is given by
\begin{align}
H_{0}  &  =t_{x/y/z}\sum_{i,a}(-1)^{a}\left(  \hat{c}_{i,a}^{\dagger}\hat
{c}_{i+\bar{\delta}_{1/2/3},a}+h.c.\right) \nonumber \\
&  -2t_{xy}\left(  1+\cos k_{0}\right)  \sum_{i,a}(-1)^{a}\hat{c}%
_{i,a}^{\dagger}\hat{c}_{i,a}-2t_{z0}\sum_{i,a}(-1)^{a}\hat{c}_{i,a}^{\dagger
}\hat{c}_{i,a}\nonumber \\
&  +t_{xz}^{\prime}\sum_{i}(e^{-ik_{x0}}\hat{c}_{i,1}^{\dagger}\hat{c}%
_{i+\bar{b}_{1},2}+e^{ik_{x0}}\hat{c}_{i,1}^{\dagger}\hat{c}_{i-\bar{b}_{1}%
,2}\nonumber \\
&  -e^{-ik_{x0}}\hat{c}_{i,1}^{\dagger}\hat{c}_{i+\bar{b}_{2},2}-e^{ik_{x0}%
}\hat{c}_{i,1}^{\dagger}\hat{c}_{i-\bar{b}_{2},2}+h.c.)\nonumber \\
&  +t_{yz}^{\prime}\sum_{i}(-ie^{-ik_{y0}}\hat{c}_{i,1}^{\dagger}\hat
{c}_{i+\bar{b}_{3},2}-ie^{ik_{y0}}\hat{c}_{i,1}^{\dagger}\hat{c}_{i-\bar
{b}_{3},2}\nonumber \\
&  +ie^{-ik_{y0}}\hat{c}_{i,1}^{\dagger}\hat{c}_{i+\bar{b}_{4},2}+ie^{ik_{y0}%
}\hat{c}_{i,1}^{\dagger}\hat{c}_{i-\bar{b}_{4},2}+h.c.) \label{ham}%
\end{align}
where $a=1,2$ is the orbital degree of freedom. $\hat{c}_{i,a}$ is the
annihilation operator of the electron at the site $i$ with orbital degree of
freedom. $t_{x/y/z}$ are the nearest neighbor hoppings in $x/y/z$ direction,
$t_{xz}^{\prime}/t_{yz}^{\prime}$ are the orbital-flip hoppings in $xoz/yoz$
plane. $t_{xy}/t_{z0}$ are the effective Zeeman field. $k_{0}$ determines the
radius of the nodal line. $k_{x0}=0.4\pi,k_{y0}=0.4\pi$ are to eliminate the
Weyl points. $\bar{\delta}_{1/2/3}$\ are the nearest vectors which are
$\left(  a_{0},0,0\right)  ,$ $\left(  0,a_{0},0\right)  ,$ $\left(
0,0,a_{0}\right)  ,$ $\bar{b}_{1/2/3/4}$ are the next nearest vectors which
are $\left(  a_{0},0,a_{0}\right)  ,$ $\left(  a_{0},0,-a_{0}\right)  ,$
$\left(  0,a_{0},a_{0}\right)  ,$ $\left(  0,a_{0},-a_{0}\right)  .$ The
lattice constant $a_{0}$ is set to be unit. It is obvious that not only the
inversion symmetry but also the time-reversal symmetry are broken.


Using Fourier transformation, we obtain the Hamiltonian in momentum space
\begin{equation}
H_{0}=\sum_{k}C_{k}^{\dagger}\mathcal{H}\left(  \mathbf{k}\right)  C_{k}%
\end{equation}
with
\begin{equation}
\mathcal{H}\left(  \mathbf{k}\right)  =\vec{h}\left(  \mathbf{k}\right)
\cdot \hat{\sigma} \label{H1}%
\end{equation}
where $C_{k}^{\dagger}=\left(  C_{k,1}^{\dagger},C_{k,2}^{\dagger}\right)  $,
$\hat{\sigma}=\left(  \sigma_{x},\sigma_{y},\sigma_{z}\right)  $ is the Pauli
matrix, and $\vec{h}\left(  \mathbf{k}\right)  =\left(  h_{x}\left(  k\right)
,h_{y}\left(  k\right)  ,h_{z}\left(  k\right)  \right)  $ with
\begin{align*}
h_{x}\left(  k\right)   &  =-4t_{xz}^{\prime}\sin \left(  k_{x}-k_{x0}\right)
\sin \left(  k_{z}\right) \\
h_{y}\left(  k\right)   &  =-4t_{yz}^{\prime}\sin \left(  k_{y}-k_{y0}\right)
\sin \left(  k_{z}\right) \\
h_{z}\left(  k\right)   &  =-2t_{x}\cos k_{x}-2t_{y}\cos k_{y}-2t_{z}\cos
k_{z}\\
&  +2t_{x/y}\left(  1+\cos k_{0}\right)  +2t_{z0}%
\end{align*}
Then, the spectrum for free fermions is derived as
\begin{equation}
E_{\mathbf{k,}\pm}=\pm \sqrt{h_{x}^{2}\left(  k\right)  +h_{y}^{2}\left(
k\right)  +h_{z}^{2}\left(  k\right)  }%
\end{equation}
In the following parts of the paper, the hopping parameters are set to
$t_{x}=t_{y}=t_{z}=t_{xy}=t_{z0}=t$.

\begin{figure}[ptbh]
\includegraphics[width = 8cm]{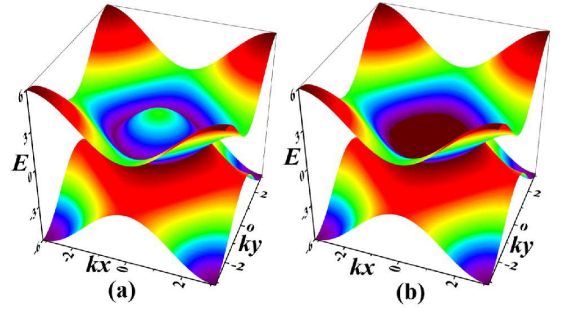}\caption{(Color online) (a) The
illustration of the nodal line locates at $k_{x}$-$k_{y}$ plane for $k_{z}=0$
with the radius of $k_{0}=\pi/2$. (b) The surface states with periodic
boundary conditions along $x$ and $y$-direction but open boundary condition
along $z$-direction. There is a drumhead inside the nodal line. The parameters
are $k_{0}=\pi/2,t_{xz}^{\prime}=t_{yz}^{\prime}=0.5t.$}%
\label{pedge}%
\end{figure}

Next, we study the nodal line of the nodal line semimetal. In $k_{x}$-$k_{y}$
plane, the nodal line satisfy the equation of
\begin{equation}
\cos k_{x}+\cos k_{y}=1+\cos k_{0} \label{eq1}%
\end{equation}
FIG.\ref{pedge}(a) shows the spectrum at $k_{z}=0$. For this case, the nodal
line locates at $k_{x}$-$k_{y}$ plane with the the radius of $k_{0}=\pi/2$,
$t_{xz}^{\prime}=t_{yz}^{\prime}=0.5t$.

Additionally, we study the surface state of the nodal line semimetal. We
consider a system with periodic boundary conditions (PBC) along $x$ and $y$
direction, but open boundary conditions (OBC) along $z$ direction. By
numerical calculations, the surface states are obtained in FIG.\ref{pedge}(b).
Comparing with FIG.\ref{pedge}(a), one can see that there exists a drumhead
induced by the nodal line which is consistent with previous
articles\cite{Yurui,NL}, and the fermi surface like a disk in the core of the BZ.

\section{Type II nodal line semimetal}

\begin{figure}[ptbh]
\includegraphics[width = 8cm]{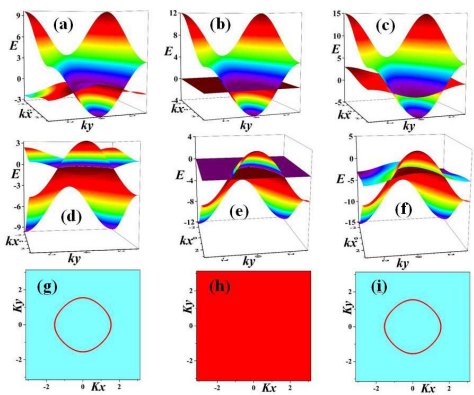}\caption{(Color online) The dispersion
of bulk states of type-II nodal line SM. $\left(  a\right)  $-$\left(
c\right)  $ are $C=0.6,C=1.0$ and $C=1.5$; $\left(  d\right)  $-$\left(
f\right)  $ are $C=-0.6,C=-1.0$ and $C=-1.5$. The tilting of the spectrums
towards to\ the center of the nodal line for the case of $C>0;$ while away
from the center of the nodal line for the case of $C<0.$ There is a flat band
Fermi surface when $\left \vert C\right \vert =1$ in $\left(  b\right)  $ and
$\left(  e\right)  .$ For the case of $\left \vert C\right \vert >1,$ one of the
energy bands reverses. $\left(  g\right)  $-$\left(  i\right)  $ are Fermion
surface of the bulk system of $\left \vert C\right \vert =0.6, 1.0$ and $1.5$.}%
\label{fig5}%
\end{figure}

In this part, a new type of NLSM named type-II NLSM is proposed. To get a
typical type-II NLSM, we add a new term into the original model as
\begin{equation}
\mathcal{H}\left(  \mathbf{k}\right)  =\vec{h}\left(  \mathbf{k}\right)
\cdot \hat{\sigma}+h_{0}I
\end{equation}
with
\begin{align}
h_{0}\left(  k\right)   &  =C[-2t_{x}\cos \left(  k_{x}\right)  -2t_{y}%
\cos \left(  k_{y}\right) \nonumber \\
&  +2t_{x/y}\left(  1+\cos \left(  k_{0}\right)  \right)  ]
\end{align}
$C$ is a coefficient that determines the type of a NLSM. $\left \vert
C\right \vert =1$ is a critical point: when $\left \vert C\right \vert <1$, the
NLSM belongs to type-I nodal line SM; when $\left \vert C\right \vert >1$, the
NLSM belongs to type-II nodal line SM. At the critical point $\left \vert
C\right \vert =1$, NLSM has a flat band at Fermi surface as FIG.\ref{fig5}
$\left(  h\right)  $ (the red region). For the case of $\left \vert
C\right \vert >1$, one of the energy bands reverses.

The sign of coefficient $C$ denotes the tilting orientation: When $C>0$, the
tilting of the spectra towards to the center of the node-line, while away from
the center when $C<0$. Numerical calculation of dispersions is shown in
FIG.\ref{fig5}: the coefficient $C>0$ for $\left(  a\right)  $-$\left(
c\right)  $, while $C<0$ for $\left(  d\right)  $-$\left(  f\right)  .$ We can
see clearly that the tilting of the nodal line towards to the center of the
nodal line when $C>0;$ while away from the center when $C<0$, $\left(
g\right)  $-$\left(  i\right)  $ are Fermion surface of the bulk system for
$\left \vert C\right \vert =0.6, 1.0$ and $1.5$.

\begin{figure}[ptbh]
\includegraphics[width = 8cm]{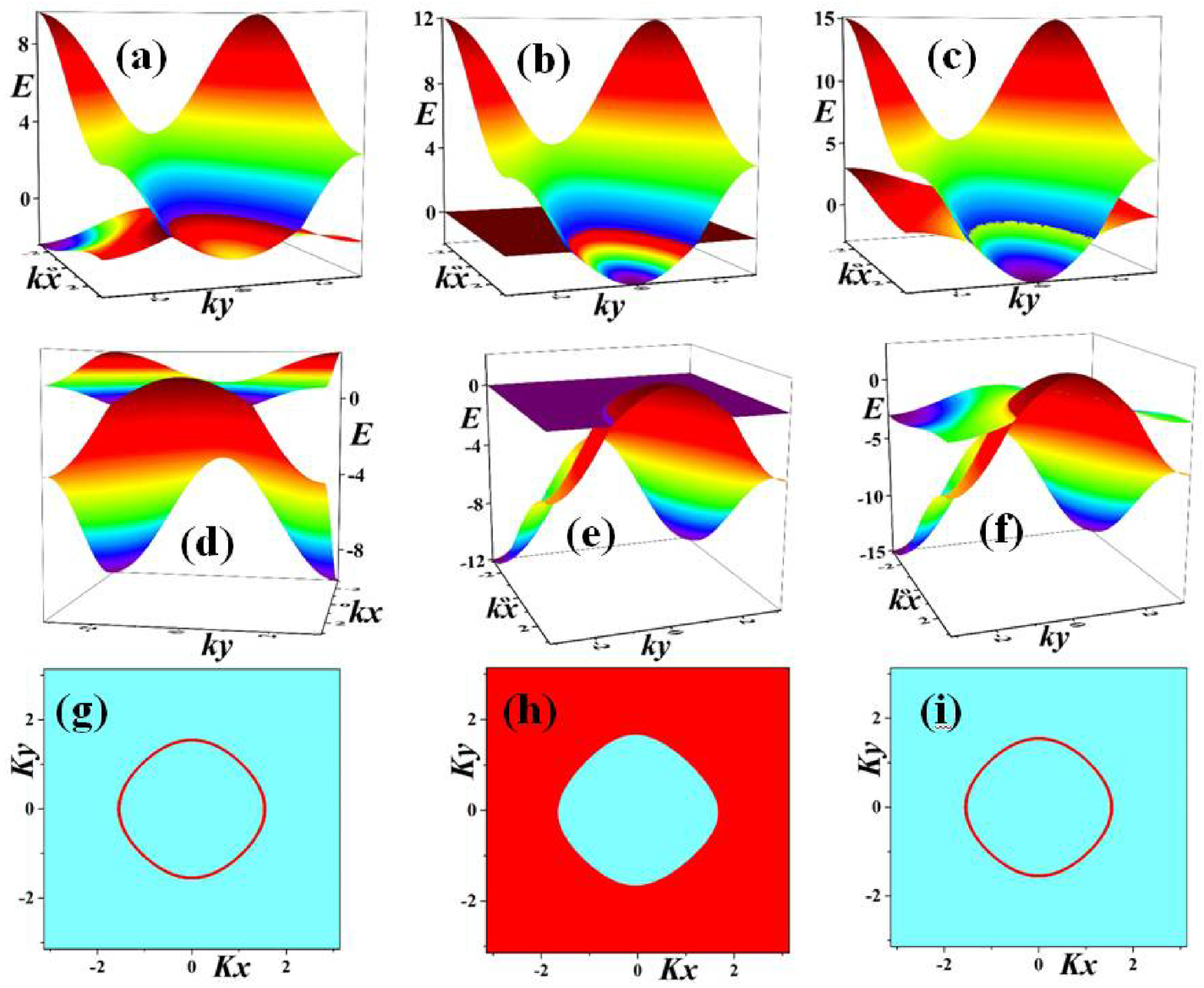}\caption{(Color online) The dispersion
of surface states of type-II nodal line SM. $\left(  a\right)  $-$\left(
c\right)  $ are $C=0.6,C=1.0$ and $C=1.5$; $\left(  d\right)  $-$\left(
f\right)  $ are $C=-0.6,C=-1.0$ and $C=-1.5$. Due to the existence of the
tilting term of $C\neq0,$ the drumhead-like surface flat bands like
FIG.\ref{pedge} (b) disappears. For the case of $\left \vert C\right \vert <1$
($\left(  a\right)  $ and $\left(  d\right)  $)$,$ the Fermi surface is a
circle (like $\left(  g\right) $); at the critical point $\left \vert
C\right \vert =1$ ($\left(  b\right)  $ and $\left(  e\right)  $)$,$ the Fermi
surface changes into a flat band with a hole in the center (like $\left(
h\right) $); for the case of $\left \vert C\right \vert >1$ ($\left(  c\right)
$ and $\left(  f\right)  $)$,$ the Fermi surface changes back into a circle
one (like $\left( i\right) $).}%
\label{fig6}%
\end{figure}

We then study the topological properties of Type-II nodal line SM. The
topological protected surface state is a hallmark of topological system. In
type-II nodal line semimetal, the surface states show similar behavior of the
nodal states on bulk system -- the surface states can also tilted and becomes
`type-II'. In tilted NLSM, the surface states are shown in FIG.\ref{fig6}
which are top views from $z$ axis for lowest two bands near Fermi surface. In
FIG.\ref{fig6}, the coefficient $C>0$ for $\left(  a\right)  $-$\left(
c\right)  $, while $C<0$ for $\left(  d\right)  $-$\left(  f\right)  $. Due to
the tilting effect for the case of $C\neq0,$ the drumhead-like surface flat
band like FIG.\ref{pedge} $(b)$ disappears and instead by a dispersive one.
Thus, the surface states in NLSM can also be tilted like nodal line in bulk,
which is similar as type-II Weyl semimetal\cite{WSM2}.

We discuss the evolution of Fermi surface of lowest energy band of bulk
states. In type-I NLSM with $C=0$, the Fermi surface of bulk states is a
circle at $\mu=0$ (here $\mu$ is the chemical potential). At the critical
point $\left \vert C\right \vert =1$, one band of NLSM becomes flat, which leads
to a tilted surface state. While the Fermi surface of surface states is a disk
at $\mu=0$ when $C=0$. At the critical point $\left \vert C\right \vert =1$, it
becomes a flat band with a hole in the center like FIG.\ref{fig6} $(h)$.

\begin{figure}[ptbh]
\includegraphics[width = 8cm]{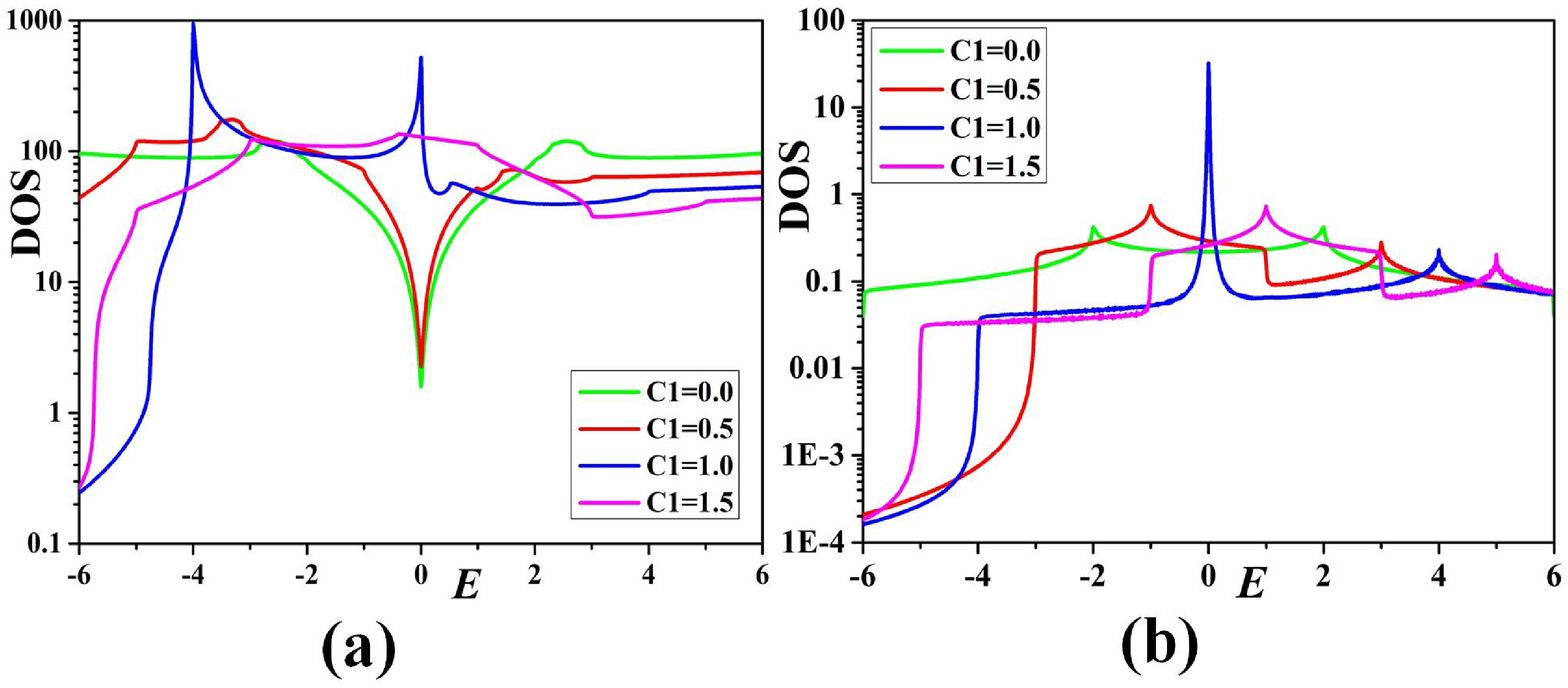}\caption{(Color online) The DOS for
Type-II nodal line SM. $\left(  a\right)  $ for the whole Brillouin zone;
$\left(  b\right)  $ The DOS for $k_{z}=0$ where the nodal line locates.}%
\label{fig7}%
\end{figure}

In addition, we also calculate the density of states (DOS). The expression for
calculating DOS is
\[
\rho \left(  \omega \right)  =-\frac{1}{\pi}\operatorname{Im}\sum_{\sigma
,k}G_{\sigma}\left(  \omega,k\right)
\]
where $G_{\sigma}\left(  \omega,k\right)  $ is Matsubara Green's Function
which are
\begin{align}
G_{\uparrow}\left(  \omega,k\right)   &  =\frac{\left \vert E_{\mathbf{k,}\pm
}\right \vert +h_{z}}{2\left \vert E_{\mathbf{k,}\pm}\right \vert }\frac
{1}{\omega+i\eta-\left(  h_{0}+E_{\mathbf{k,}+}\right)  }\nonumber \\
&  +\frac{\left \vert E_{\mathbf{k,}\pm}\right \vert -h_{z}}{2\left \vert
E_{\mathbf{k,}\pm}\right \vert }\frac{1}{\omega+i\eta-\left(  h_{0}%
+E_{\mathbf{k,}-}\right)  },\\
G_{\downarrow}\left(  \omega,k\right)   &  =\frac{\left \vert E_{\mathbf{k,}%
\pm}\right \vert -h_{z}}{2\left \vert E_{\mathbf{k,}\pm}\right \vert }\frac
{1}{\omega+i\eta-\left(  h_{0}+E_{\mathbf{k,}+}\right)  }\nonumber \\
&  +\frac{\left \vert E_{\mathbf{k,}\pm}\right \vert +h_{z}}{2\left \vert
E_{\mathbf{k,}\pm}\right \vert }\frac{1}{\omega+i\eta-\left(  h_{0}%
+E_{\mathbf{k,}-}\right)  }.
\end{align}
Here $\eta$ is an infinite small quantity and real, $\omega$ is the energy
level. After considering the tilting effect on the spectra, the DOS changes
correspondingly. In FIG.\ref{fig7}(a) there always exists a sharp peak at
$E=0$ due to the flat band states for type-II NLSM. In FIG.\ref{fig7}(b), for
$k_{z}=0,$ owing to the existence of bulk flat band, there exists a sharp peak
at $E=0$ for the case of $\left \vert C\right \vert =1.$

\section{Effect of magnetic field on type-II nodal line semimetal}

In type-II Weyl semimetal, the negative magnetic effect (NME) becomes
anisotropic. The failure of NME in the prescribed direction is caused by the
collapsion of Landau level\cite{Yao}. We now show that the collapsion of
Landau level also appears in nodal line semimetal.

We add the magnetic field along $x$-direction, i.e., $\mathbf{B}=B\hat{x}$ and
$\mathbf{A}=\left(  0,Bz/2,-By/2\right)  ,$ then use the usual Peierls
substitutions $k_{x}\rightarrow \tilde{k}_{z}-eBy/2$, $k_{y}\rightarrow
\tilde{k}_{y}+eBz/2$. We introduce the ladder operators
\begin{align*}
a^{\dag}  &  =\left[  \hbar \partial_{y}+i\hbar \partial_{z}-\left(  \frac
{eBy}{2}+i\frac{eBz}{2}\right)  \right]  ,\\
a  &  =\left[  -\left(  \hbar \partial_{y}-i\hbar \partial_{z}\right)  -\left(
\frac{eBy}{2}-i\frac{eBz}{2}\right)  \right]  ,
\end{align*}
where $\tilde{k}_{x}=-i\hbar \partial_{x}$, $\tilde{k}_{y}=-i\hbar \partial_{y}%
$. These operators rise and fall the Landau levels of free electrons as
\begin{equation}
a^{\dag}|n,k_{x}\rangle=\sqrt{n+1}|n+1,k_{x}\rangle
\end{equation}
and
\begin{equation}
a|n,k_{x}\rangle=\sqrt{n}|n-1,k_{x}\rangle,
\end{equation}
where $|n,k_{x}\rangle$ is the free electrons Landau level wave-function. When
an electron occupies the state $|n,k_{x}\rangle$, it rounds in circles in
$y$-$z$ plane. The translation invariance along $x$-direction is preserved so
that $k_{x}$ is still a good quantum number.

We expanse Hamiltonian near nodal line and only keep first-order terms,
considering the perturbation along radial $\left(  \Delta k_{R}\right)  $ and
tangential $\left(  \Delta k_{T}\right)  $ directions of the nodal line. After
a unitary transformation between two coordinates, we have
\[
\Delta k_{x}\sigma_{x}+\Delta k_{y}\sigma_{y}=\Delta k_{T}\sigma_{T}+\Delta
k_{R}\sigma_{R}%
\]
where $\sigma_{T}=\sigma_{x}\sin \theta-\sigma_{y}\cos \theta,$ $\sigma
_{R}=\sigma_{x}\cos \theta+\sigma_{y}\sin \theta$ and $k_{T}=k_{x}\sin
\theta-k_{y}\cos \theta,$ $k_{R}=k_{x}\cos \theta+k_{y}\sin \theta$, and $\theta$
is the intersection angle with $x$-axis in $x$-$y$ plane. Then, the
Hamiltonian variation induced by the perturbation is
\begin{align}
\Delta \mathcal{H}\left(  \mathbf{k}\right)   &  =-2\Delta k_{z}\left(
k_{0}\sigma_{R}+\Delta k_{T}\sigma_{T}+\Delta k_{R}\sigma_{R}\right)
\nonumber \\
&  +\left(  2k_{0}\Delta k_{R}+\Delta k_{z}^{2}\right)  \sigma_{z}%
+2Ck_{0}\Delta k_{R}\sigma_{0}\nonumber \\
&  \simeq-2k_{0}\Delta k_{z}\sigma_{R}+2k_{0}\Delta k_{R}\sigma_{z}%
+2Ck_{0}\Delta k_{R}\sigma_{0}, \label{LL2}%
\end{align}
which is independent of $k_{T}$ because of there is no dispersion along nodal
line. As magnetic field is applied along $x$-direction, and $A_{T}$
(tangential directions of $\mathbf{A}$) is irrelevant, we focus tangential
component of magnetic field $B\sin \theta$. The corresponding Landau levels
near the nodal line becomes
\begin{align}
E_{n\geq1}  &  =\pm v_{0}\sqrt{2n\alpha^{3}e\hbar B\sqrt{1-(2k_{x}/\pi)^{2}}%
}\\
E_{n=0}  &  =0
\end{align}
where $v_{0}=2k_{0},$ $\alpha=\sqrt{1-\beta^{2}},$ $\beta=C,$ $e$ is
elementary charge, $\hbar$ is Planck constant. In type-I region, the zeroth
level $E=0$ is maintained; in type-II region $\left \vert C\right \vert >1$,
$1-\beta^{2}<0$, so that $\alpha$ is imaginary and the expression is invalid.
This corresponds to collapsing of Landau levels mentioned in Ref. \cite{Yao}.
The zeroth Landau level also disappears.

\begin{figure}[ptbh]
\includegraphics[width = 8cm]{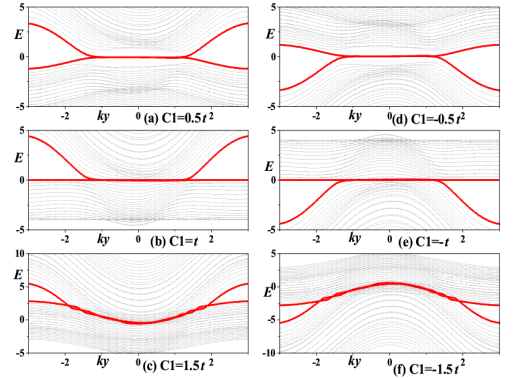}\caption{(Color online) Dispersion of
type-II NLSM in a magnetic field along x-direction with finite tilting
strength }%
\label{fig9}%
\end{figure}

In FIG.\ref{fig9}, we also give the numerical results with different tilting
strengthes $C$. There are two flat bands near nodal line when $\left \vert
C\right \vert <1$, which correspond to zeroth Landau level. When $\left \vert
C\right \vert >1$, the flat bands disappears, and the system becomes metal
which is similar to Weyl semimetal\cite{WSM2,WSM3,kong}.

\section{Correlation effect on type-II nodal line semimetal}

In this part, we study the correlation effect on type-II NLSM by
considering an on-site repulsive interaction. Then the Hamiltonian is
rewritten as
\begin{equation}
H=H_{0,\uparrow}+H_{0,\downarrow}+U\sum_{i,a}\hat{n}_{i,\uparrow,a}\hat
{n}_{i,\downarrow,a}-\mu \sum_{i,\tau,a}\hat{c}_{i,\tau,a}^{\dagger}\hat
{c}_{i,\tau,a}%
\end{equation}
where $H_{0,\uparrow}$ and $H_{0,\downarrow}$ are the Hamiltonians of
Eq.(\ref{ham}) after considering the spin degree of freedom. $\hat{n}%
_{i,\tau,a}=\hat{c}_{i,\tau,a}^{\dagger}\hat{c}_{i,\tau,a}$ is the operator of
particle number with two spin degrees of freedom $\tau$ and two orbital
degrees of freedom $a$, $U$ is the on-site Coulomb repulsive interaction
strength and $\mu$ is the chemical potential.

\begin{figure}[ptbh]
\includegraphics[width = 8cm]{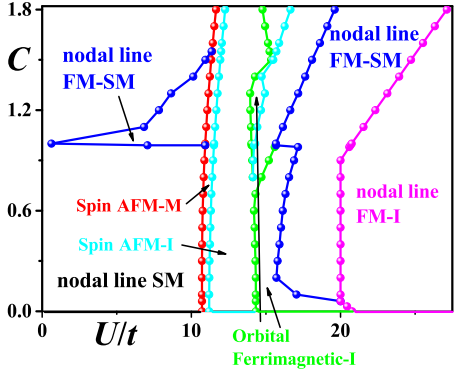}\caption{(Color online) Global phase
diagram for different tilting strengthes $C.$ There exist six phases: nodal
line SM without any magnetic order, nodal line SM with FM order of spin degree
of freedom (FM-SM), metal with AF order of spin degree of freedom (Spin
AFM-M), insulator with AF order of spin degree of freedom (Spin AFM-I),
insulator with Ferrimagnetic order of orbital degree of freedom (Orbital
Ferrimagnetic-I) and nodal line insulator with FM order of spin degree of
freedom (FM-I). In the global phase diagram, there are two kinds of quantum
phase transitions: one is the quantum phase transition between a long range
ordered state and a phase without the long range order, the other is
metal-insulator transition that is characterized by the condition of zero
fermion's energy gaps.}%
\label{fig10}%
\end{figure}

Because the orbital SU(2) rotation symmetry is broken, when considering the
repulsive interaction, magnetic order of spin degree of freedom may appears
and the corresponding spin SU(2) rotation symmetry is spontaneously broken. By
the mean field theory, the ferromagnetic (FM) order of spin degree of freedom
for bulk states is denoted by
\begin{equation}
\left \langle n_{i,\tau}\right \rangle =\frac{1}{2}\left(  n+\tau M_{F}\right)
\label{order}%
\end{equation}
where $n$ is the number of particles, and we only consider the half-filling
case for $n=1.$ $\tau=1$ represents spin up and $\tau=-1$ represents spin
down. $M_{F}$ is the FM order parameter of spin degree of freedom. We can
write the self-consistent equations as%
\begin{align}
\left \langle n_{i\uparrow}\right \rangle +\left \langle n_{i\downarrow
}\right \rangle  &  =1,\\
\left \langle n_{i\uparrow}\right \rangle -\left \langle n_{i\downarrow
}\right \rangle  &  =M_{F}.
\end{align}

After Fourier transformation, the self-consistent equations in momentum space
can be rewritten as
\begin{align}
M_{F}  &  =\frac{1}{2N}\sum_{k}\left[  \theta \left(  -E_{1\uparrow}\right)
+\theta \left(  -E_{2\uparrow}\right)  -\theta \left(  -E_{1\downarrow}\right)
-\theta \left(  -E_{2\downarrow}\right)  \right]  ,\\
1  &  =\frac{1}{2N}\sum_{k}\left[  \theta \left(  -E_{1\uparrow}\right)
+\theta \left(  -E_{2\uparrow}\right)  +\theta \left(  -E_{1\downarrow}\right)
+\theta \left(  -E_{2\downarrow}\right)  \right]  ,
\end{align}
where $\theta \left(  x\right)  $ is a step-up function and $\theta \left(
x\right)  =1$ for $x>0$ and $\theta \left(  x\right)  =0$ for $x<0$, $N$ is the
number of the unit cells and
\begin{align*}
E_{1\uparrow}  &  =h_{0}-\frac{UM_{F}}{2}-\mu_{eff}-E_{\mathbf{k}};\\
E_{2\uparrow}  &  =h_{0}-\frac{UM_{F}}{2}-\mu_{eff}+E_{\mathbf{k}};\\
E_{1\downarrow}  &  =h_{0}+\frac{UM_{F}}{2}-\mu_{eff}-E_{\mathbf{k}};\\
E_{2\downarrow}  &  =h_{0}+\frac{UM_{F}}{2}-\mu_{eff}+E_{\mathbf{k}},
\end{align*}
with $\mu_{eff}=\mu-\frac{U}{2}.$

At the mean field level, we can also define other long range orders: the
antiferromagnetic (AF) order of spin degree of freedom for bulk states
\begin{equation}
\left \langle n_{i,\tau}\right \rangle =\frac{1}{2}\left[  n+(-1)^{i}\tau
M_{AF}\right]
\end{equation}
where $M_{AF}$ is the AF order parameter of spin degree of freedom; the
ferromagnetic (FM) order of orbital degree of freedom for bulk states
\begin{equation}
\left \langle n_{i,a}\right \rangle =\frac{1}{2}\left[  n+(-1)^{a}M_{F}^{\prime
}\right]
\end{equation}
where $M_{F}^{\prime}$ is FM order parameter of orbital degree of freedom;
the antiferromagnet (AF) order of orbital degree of freedom for bulk states
\begin{equation}
\left \langle n_{i,a}\right \rangle =\frac{1}{2}\left[  n+(-1)^{i}(-1)^{a}%
M_{AF}^{\prime}\right]
\end{equation}
where $M_{AF}^{\prime}$ is AF order parameter of orbital degree of freedom.
These numerical calculations are the same as the FM case of spin degree of freedom.

\begin{figure}[ptbh]
\includegraphics[width = 0.55\textwidth]{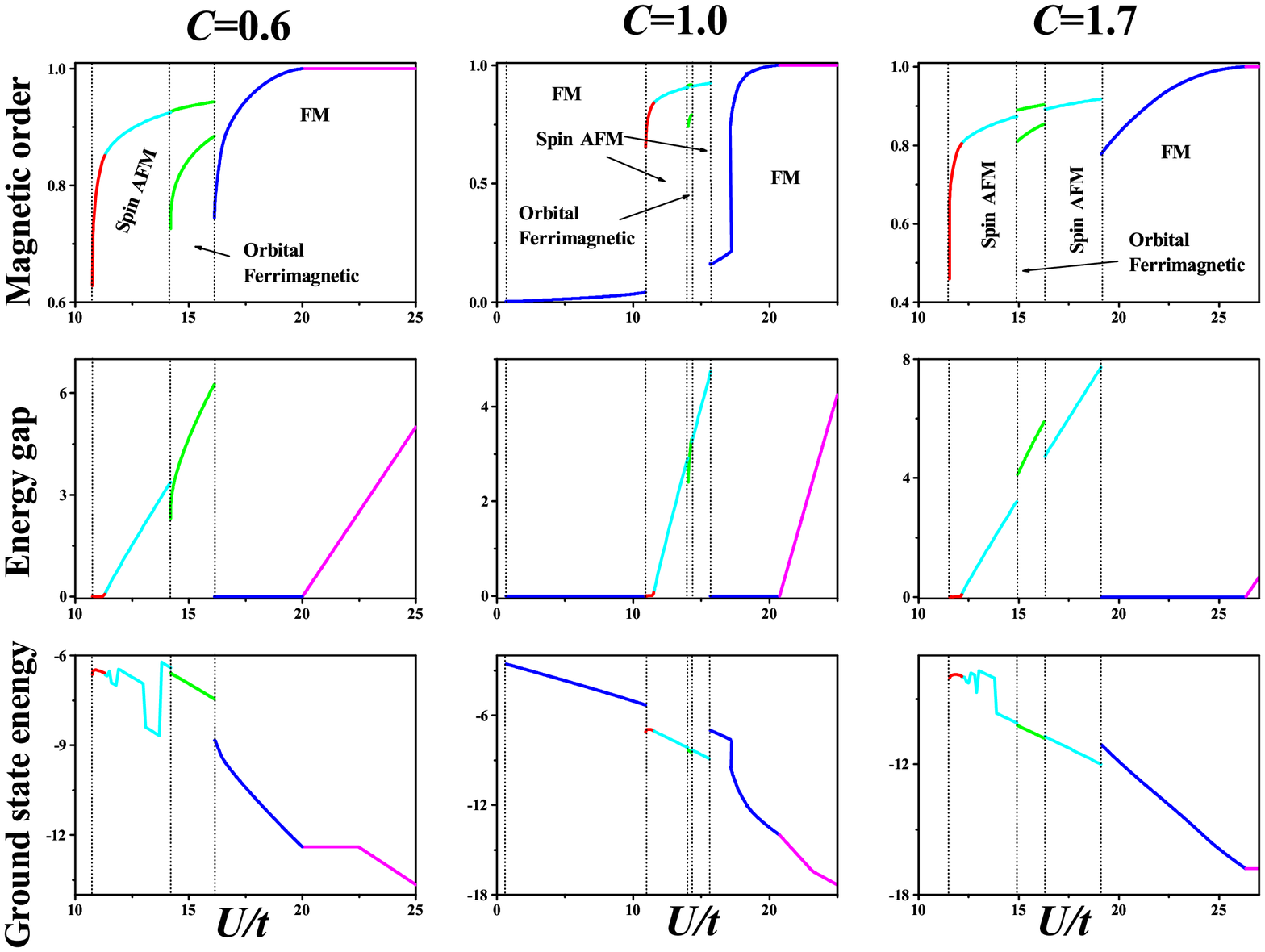}\caption{(color online) The
first, second and third rows represent magnetization, the energy gap and the
ground state energy respectively. Different columns represent different
tilting strengthes. In these figures, we use different colored lines represent
different phases, like blue line represents nodal line SM-FM, red line
represents Spin AFM-M, cyan line represents Spin AFM-I, green line represents
Orbital Ferrimagnetic-I and magenta line represents nodal line FM-I. We use
black dotted lines to distinguish different magnetic order phases.}%
\label{fig11}%
\end{figure}

Then by using mean field approach, we obtain the global phase diagram for
different NLSMs with different tilting strengthes $C$ in FIG.\ref{fig10}. In
FIG.\ref{fig10}, there exist six phases: nodal line SM without any long range
order, nodal line SM with FM order of spin degree of freedom (FM-SM), metal
with AF order of spin degree of freedom (Spin AFM-M), insulator with AF order
of spin degree of freedom (Spin AFM-I), insulator with Ferrimagnetic order of
orbital degree of freedom (Orbital Ferrimagnetic-I) and nodal line insulator
with FM order of spin degree of freedom (FM-I). In the global phase diagram,
there are two kinds of quantum phase transitions: one is the quantum phase
transition between a long range ordered state and a phase without the long
range order, the other is metal-insulator transition that is characterized by
the condition of zero fermion's energy gaps.

In the global phase diagram, a remarkable result is about the magnetic phase
transition at $C=1.$ For the case of $C =1$, there exists a flat band Fermi
surface (See FIG.\ref{fig5}(h)). As a result, a very tiny repulsive
interaction will induce an FM order of spin degree of freedom (See the result
in FIG.\ref{fig10}). In FIG.\ref{fig11}, we also plot the magnetization, the
energy gap and the ground state energy via the repulsive interaction for the
cases $C=0.6$, $C=1.0$ and $C=1.7,$ respectively. The first, second and third
rows represent magnetization, the energy gap and the ground state energy
respectively. Different columns represent different tilting strengthes. In
these figures, we use different colored lines represent different phases, like
blue line represents nodal line SM-FM, red line represents Spin AFM-M, cyan
line represents Spin AFM-I, green line represents Orbital Ferrimagnetic-I and
magenta line represents nodal line FM-I. We use black dotted
lines to distinguish different magnetic order phases.

\begin{figure}[ptbh]
\includegraphics[width = 8cm]{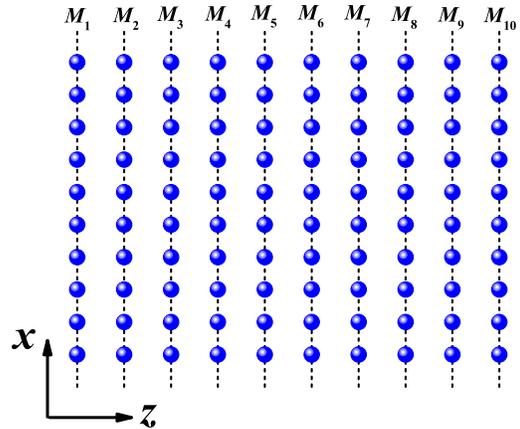}
\caption{(Color online) The illustration of the OBC for correlated effect on
surface states. The system with periodic boundary conditions along x and
y-direction, but open boundary conditions along z-direction. $M_{1}$-$M_{10}$
are ferromagnetic orders for different levels of system.}%
\label{tushi}%
\end{figure}

Next, we consider the correlated effect on surface states and show the
interaction-induced surface orders in the NLSMs. Because the orbital SU(2)
rotation symmetry is broken and the antiferromagnetic order of spin degree of
freedom for surface states is not well defined, we focus on ferromagnetic
order of spin degree of freedom for surface states.

\begin{figure}[ptb]
\scalebox{0.4}{\includegraphics* [0.5in,0.0in][9.5in,10.5in]{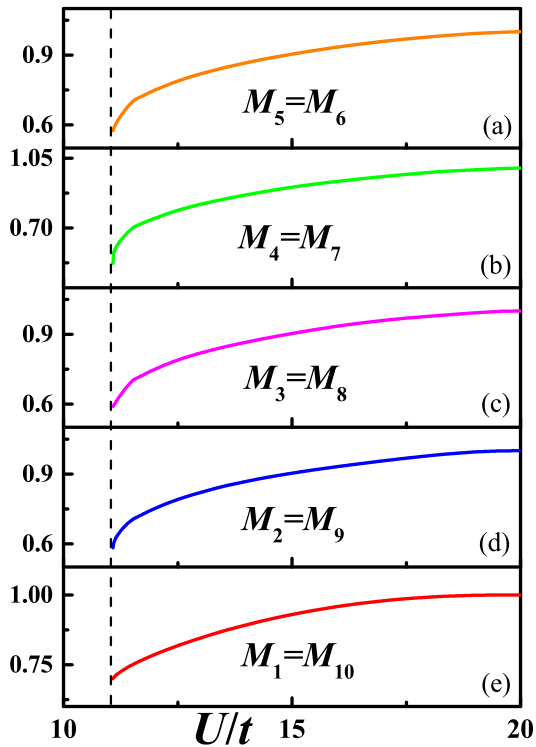}}
\caption{(color online) The ferromagnetic order $M$ of the system with open
boundary condition for the case of $C=0.1$. The black dash line represents the
magnetic phase transition from $M=0$ to $M\neq0$. $(a)$-$(e)$ are the FM
orders of different sites. One can see that the surface FM order is more robust than
the bulk FM order. }%
\label{eorder}%
\end{figure}

\begin{figure}[ptb]
\scalebox{0.32}{\includegraphics* [0in,0.1in][9.5in,9in]{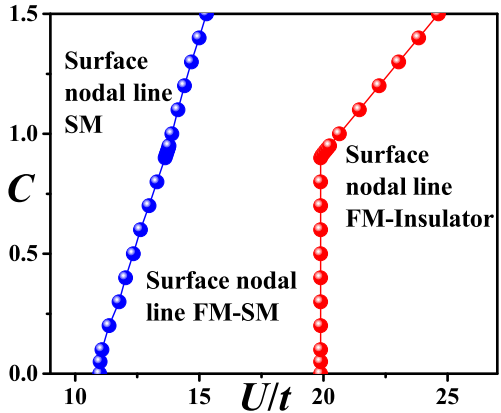}}
\caption{(color online) Global phase
diagram for different tilting strengthes $C$ on surface states of NLSMs. There
are three phases: surface nodal line SM without any magnetic order, surface
nodal line SM with FM order of spin degree of freedom (FM-SM), and surface
nodal line insulator with FM order of spin degree of freedom (FM-insulator).
There are two phase transition: the magnetic phase transition and the
metal-insulator phase transition. }%
\label{fig13}%
\end{figure}

Because the nodal line locates at $k_{x}$-$k_{y}$ plane, we consider a system
with periodic boundary conditions (PBC) along $x$ and $y$-direction, but open
boundary conditions (OBC) along $z$-direction. Now, due to SU(2) spin rotation
symmetry, the ansatz of FM order of spin degree of freedom is the same as
Eq.(\ref{order}). Along $z$-direction, the system have $10$ lattice site like
Fig.\ref{tushi}. Because there is no translation symmetry along
$z$-direction, we must calculate the mean field ansatz of FM order
site-by-site. After considering inverse symmetry, there are five different
cases to calculate. In Fig.\ref{eorder} $(a)$-$(e)$ are the FM orders of on
different lattice sites along $z$-direction.

After numerical calculations, we get the global phase diagram for different
types of NLSMs with OBC in FIG.\ref{fig13}. Comparing with FIG.\ref{fig10},
there exist three phases: surface nodal line SM without any magnetic order,
surface nodal line SM with FM order of spin degree of freedom (FM-SM), and
surface nodal line insulator with FM order of spin degree of freedom
(FM-insulator). There are two phase transition: the magnetic phase transition
and the metal-insulator phase transition. Due to the effect of OBC, the
results are different from FIG.\ref{fig10}. When we tune the strength of
repulsive interaction, the bulk FM order appears earlier than the surface FM
order for different types of NLSMs.

\begin{figure}[ptbh]
\centering
\includegraphics[width = 0.5\textwidth]{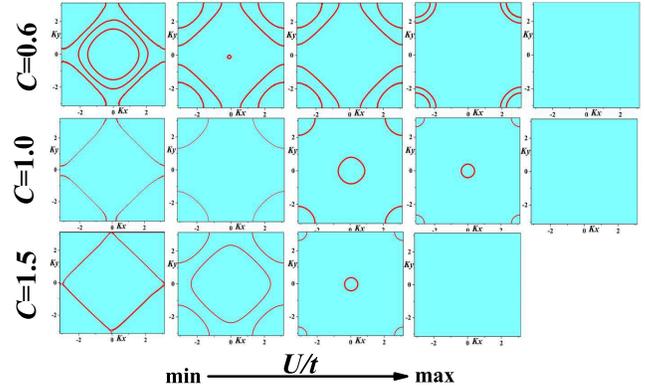}\caption{(color online)
Fermi surface for surface states with different tilting strength $C$.}%
\label{esur}%
\end{figure}

Beyond the critical tilting point $C=1$, one of the energy bands of surface
states reverses. See Fig.\ref{esur}. For different tilting strengthes, with
the increase of interaction, the shape of Fermi surface for surface states
changes, and finally the system becomes an insulator.

\section{Conclusion}

In this paper, we pointed out that there exists a new type of node-line
semimetal - type-II NLSM based on a two-band cubic lattice model. We studied
the effect of magnetic field on type-II NLSM and found the Landau level
collapsion in this system. After considering repulsive interaction and
additional spin degree of freedom, different magnetic orders appear in the
bulk states and ferromagnetic order exist in surface states. At critical point
between type-I NLSM and type-II NLSM, arbitrary tiny interaction induces
ferromagnetic order due to a flat band at Fermi surface.

Finally, we propose an experimental setup to realize the NLSM on optical
lattice. The model discussed in this paper includes complex-valued nearest and
next nearest neighbor hopping in cubic lattice. Hopefully this can be realized
in a three-dimensional optical lattice with two components of Fermi atoms such
as $^{6}$Li and $^{40}$K. The real-valued hopping can be induced by kinetic
which could be tuned by change the potential depth and the imaginary-valued
hopping could be induced by a two-photon Raman process or shaking lattice.
Similar system in one dimension and two dimensions had been realized
recently\cite{exp1,exp2}.

\begin{center}
{\textbf{* * *}}
\end{center}

This work is supported by National Basic Research Program of China (973
Program) under the grant No. 2011CB921803, 2012CB921704 and NSFC Grant No.
11174035, 11474025, 11404090, Natural Science Foundation of Hebei Province
(Grant No. A2015205189), Hebei Education Department Natural Science Foundation
(Grant No. QN2014022), SRFDP.


\begin{thebibliography}{99}                                                                                               %


\bibitem {DSM1}Z. Wang, Y. Sun, X. Q. Chen, C. Franchini, G. Xu, H. Weng, X.
Dai and Z. Fang, Phys. Rev. B 85, 195320 (2012).

\bibitem {DSM2}Z. Wang, H. Weng, Q. Wu, X. Dai and Z. Fang, Phys. Rev. B 88,
125427 (2013).

\bibitem {WSM1}H. B. Nielsen and M. Ninomiya, Phys. Lett. 130B, 389 (1983).

\bibitem {XWan}X. Wan, A. M. Turner, A. Vishwanath, and S. Y. Savrasov, Phys.
Rev. B 83, 205101 (2011).

\bibitem {SM}G. Xu, H. Weng, Z. Wang, X. Dai and Z. Fang, Phys. Rev. Lett.
107, 186806 (2011).

\bibitem {SM1}L. Balents, Physics 4, 36 (2011).

\bibitem {NLSM1}A. A. Burkov, M. D. Hook and L. Balents, Phys. Rev. B 84,
235126 (2011).

\bibitem {NLSM2}H. Weng, Y. Liang, Q. Xu, R. Yu, Z. Fang, X. Dai and Y.
Kawazoe, Phys. Rev. B 92, 045108 (2015).

\bibitem {NLSM3}L. K, Lim and R. Moessner, Phys. Rev. Lett. 118, 016401 (2017).

\bibitem {CPB}C. Fang, H. M. Weng, X. Dai and Z. Fang, Chin. Phys. B 11,
117106 (2016).

\bibitem {Lv1}B.Q. Lv, H. M. Weng, B. B. Fu, X. P. Wang,H.Miao, J. Ma, P.
Richard, X. C. Huang, L. X. Zhao, G. F. Chen, Z. Fang, X. Dai, T. Qian and H.
Ding, Phys. Rev. X 5, 031013 (2015).

\bibitem {Lv2}B. Q. Lv, N. Xu, H. M. Weng, J. Z. Ma1 P. Richard, X. C. Huang,
L. X. Zhao, G. F. Chen, C. E. Matt, F. Bisti, V. N. Strocov, J. Mesot, Z.
Fang, X. Dai, T. Qian, M. Shi and H. Ding, Nat. Phys. 11, 724 (2015).

\bibitem {Chang}T. R Chang, et al., Nat. Commun. 7, 10639 (2016).

\bibitem {Xu}N. Xu, Z. J. Wang, A. P. Weber, A. Magrez, P. Bugnon, H. Berger,
C. E. Matt, J. Z. Ma, B. B. Fu, B. Q. Lv, N. C. Plumb, M. Radovic, E.
Pomjakushina, K. Conder, T. Qian, J. H. Dil, J. Mesot, H. Ding and M. Shi,
arXiv:cond-mat/1604.02116 (2016).

\bibitem {Xie}L. S. Xie, L. M. Schoop, E. M. Seibel, Q. D. Gibson, W. Xie, and
R. J. Cava, Apl Materials 3, 083602 (2015)

\bibitem {Yurui}R. Yu, H. M. Weng, Z. Fang, X. Dai, and X. Hu, Phys. Rev.
Lett. 115, 036807 (2015).

\bibitem {WSM2}A. A. Soluyanov, D. Gresch, Z. Wang, Q. Wu, M. Troyer, X. Dai
and B. A. Bernevig, Nature (London) 527, 495 (2015).

\bibitem {WSM3}F. Y. Li, X. Luo, X. Dai, Y. Yu, F. Zhang and G, Chen, Phys.
Rev. B 94, 121105(R) (2016).

\bibitem {kong}X. Kong, J. He L. Ying and S. P. Kou, Phys. Rev. A 95, 033629 (2017).

\bibitem {aniso}Y. J. Wang, E. F. Liu, H. M. Liu, Y. M. Pan, L. Q. Zhang, J.
W. Zeng, Y. J. Fu, M. Wang, K. Xu, Z. Huang, Z. L. Wang, H. Z. Lu, D. Y. Xing,
B. G. Wang, X. G. Wan and F. Miao, Nature Communications 7, 13142 (2016).

\bibitem {Yao}Z. M. Yu and Y. G. Yao, Phys. Rev. Lett. 117, 077202 (2016).

\bibitem {NL}Youngkuk Kim, Benjamin J. Wieder, C. L. Kane, and Andrew M.
Rappe, Phys. Rev. Lett. 115, 036806 (2015).

\bibitem {exp1}Y. J. Lin, K. Jim\'{e}nez-Garc\'{\i}a and I. B. Spielman,
Nature 471, 83--86 (2011).

\bibitem {exp2}Z. Wu, L. Zhang, W. Sun, X. T. Xu, B. Z. Wang, S. C. Ji, Y. J.
Deng, S. Chen, X. J. Liu and J. W. Pan, Science 354, 6308 (2016).
\end{thebibliography}
\end{document}